\documentclass{llncs}
\usepackage{makeidx}  % allows for indexgeneration
\usepackage{llncsdoc}
\usepackage{amsmath,graphicx}
\usepackage{amssymb}
\usepackage{algorithm}
\usepackage[noend]{algpseudocode}
\usepackage{multirow}
\usepackage{array}
\usepackage[skip=3pt]{caption}
\usepackage{verbatim}
\usepackage{tabularx}
\usepackage{todonotes}
\usepackage{comment}

\begin{document}

%%%%%%%%% TITLE %%%%%%%%% 
\title{Detection and Correction of Cardiac MR Motion Artefacts during Reconstruction from k-space}
%%%%%%%%% TITLE IDEAS %%%%%%%%% 
%Detecting wrong k-space lines to correct motion artefacts using deep learning
%Using k-space based artefact detection to correct cardiac MRI motion artefacts

\newcommand*\samethanks[1][\value{footnote}]{\footnotemark[#1]}
\newcommand{\rowstyle}[1]{\gdef\currentrowstyle{#1}%
  #1\ignorespaces
}

%\begin{comment}
\author{Ilkay Oksuz$^{1}$,  
James Clough$^{1}$,  
Bram Ruijsink$^{1,2}$,  
Esther Puyol-Ant\'on$^{1}$,  
Aurelien Bustin $^{1}$,
Gastao Cruz$^{1}$,  
Claudia Prieto  $^{1}$,  
Daniel Rueckert $^{3}$,
Andrew P. King$^{1}$,
Julia A. Schnabel$^{1}$ }

\institute{$^{1}$School of Biomedical Engineering \& Imaging Sciences , King\rq{}s College London, UK \\ $^{2}$ Guy's and St Thomas' Hospital NHS Foundation Trust, London, UK \\ $^{3}$ Biomedical Image Analysis Group, Imperial College London, UK}
%\end{comment}

%\author{***** \\*****}

%\institute{*********** \\*********** \\***********}

\maketitle

%%%%%%%%% ABSTRACT
\begin{abstract}

%There has been growing interest in accelerating cardiac MR (CMR) acquisitions  using  k-space undersampling techniques. The fundamental assumption of these works is to rely on the quality of the k-space lines acquired during a pre-determined undersampled trajectory.
In fully sampled cardiac MR (CMR) acquisitions, motion can lead to corruption of k-space lines, which can result in artefacts in the reconstructed images. In this paper, we propose a method to automatically detect and correct motion-related artefacts in CMR acquisitions during reconstruction from k-space data. Our correction method is inspired by work on undersampled CMR reconstruction, and uses deep learning to optimize a data-consistency term for under-sampled k-space reconstruction. Our main methodological contribution is the addition of a detection network to classify motion-corrupted k-space lines to convert the problem of artefact correction to a problem of reconstruction using the data consistency term. We train our network to automatically correct for motion-related artefacts using synthetically corrupted cine CMR k-space data as well as uncorrupted CMR images. Using a test set of 50 2D+time cine CMR datasets from the UK Biobank, we achieve good image quality in the presence of synthetic motion artefacts. We quantitatively compare our method with a variety of techniques for recovering good image quality and showcase better performance compared to state of the art denoising techniques with a PSNR of 37.1. Moreover, we show that our method preserves the quality of uncorrupted images and therefore can be also utilized as a general image reconstruction algorithm.

%d.rueckert: The abstract reads well but I think is too long

\end{abstract}


%%%%% INTRODUCTION %%%%%

\section{Introduction}  \label{sec:intro}

%Introduction to the problem
Ensuring high image quality is essential for image analysis pipelines to extract clinically useful information. 
Misleading diagnoses can be made when the original data are of low quality, in particular for cardiac magnetic resonance (CMR) imaging, where cardiac indices are extracted using post-processing techniques including segmentation and registration. 
CMR images can contain a range of image artefacts \cite{Ferreira2013}, which can reduce the accuracy of image analysis. Improving the quality of such images acquired on MR scanners is a challenging task.

 \begin{figure}[htb]
  \centering
  \centerline{\includegraphics[width=\linewidth]{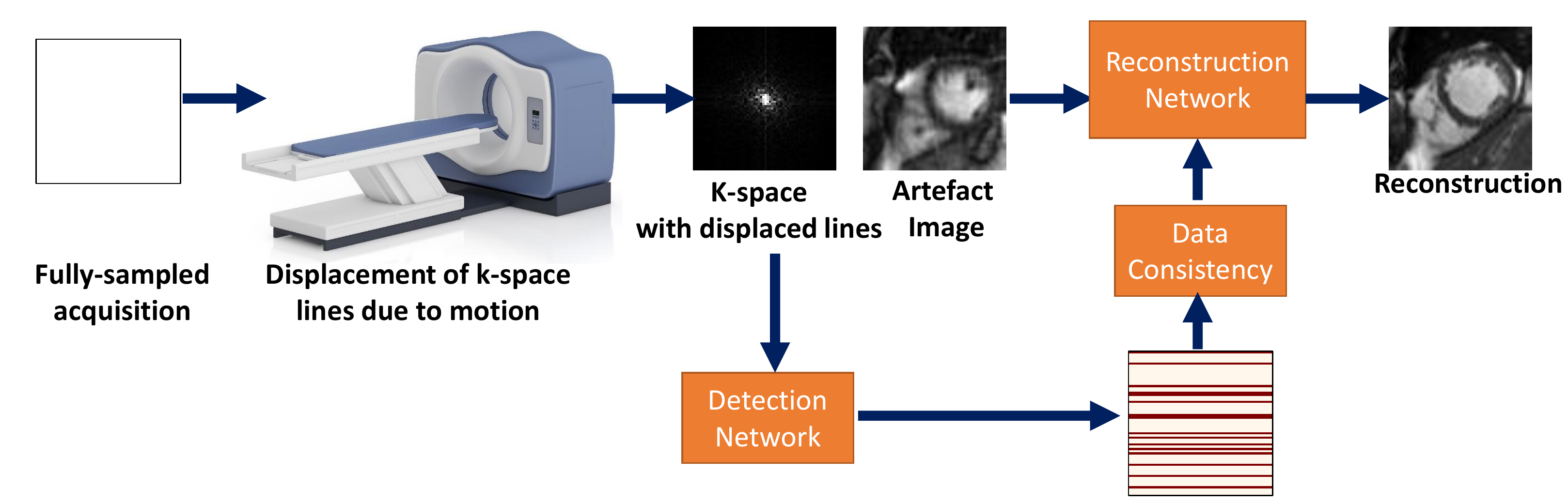}}
\caption{Detection and correction of MR artefacts using predicted data consistency masks.}
%For MR artefact correction the displaced k-space lines are not pre-determined and are identified with a detection network in our algorithm, which is trained en-to-end with the reconstruction network for both detecting the displaced k-space lines and correcting the artefacts.}
\label{fig:Motivation}
\end{figure}
%\todo{AK: In Fig 1, I would change ``mistakes during acquisition'' to something like ``displacement of k-space lines due to motion''}

% Solution to the problem and impact
One approach for correcting artefacts is image reconstruction using deep neural networks. 
~%The goal of MR reconstruction is to restore a high fidelity image from partially observed measurements. 
In deep learning based reconstruction of accelerated (i.e. undersampled) CMR, a pre-determined k-space acquisition trajectory is used and those parts of k-space that have not been sampled are estimated using an inverse problem formulation to reconstruct the image \cite{Schlemper2018}.
A different, but related problem exists in fully sampled acquisitions that have been corrupted by motion artefacts, for example due to mis-triggering or arrythmias. In these cases, the data contain correct k-space lines as well as corrupted lines, but it is unknown which are correct and which are corrupted. This problem is our focus in this paper but we draw inspiration from work on accelerated imaging in devising our solution.

% Our work
%We aim to accurately correct motion artefacts in CMR images.
We propose a k-space artefact detection network that generates an individual data consistency term for any given acquisition and converts the image artefact correction task to an undersampled image reconstruction problem, which is subsequently addressed by an algorithm developed for reconstruction of undersampled CMR acquisitions (see the illustration in Fig. \ref{fig:Motivation}). Our proposed method is evaluated using 300 cine SSFP (2D+time) CMR datasets from the UK Biobank.   

%d.rueckert: How do you deal with the missing phase information?

The major contributions of this work are as follows: First, we introduce a novel solution for the detection of artefacts in CMR images. Second, we use the output of this k-space artefact detection network to introduce a data consistency term to be used by an image reconstruction network. By training both networks end-to-end we are able to ignore motion corrupted k-space lines during the reconstruction. Finally, our algorithm is trained and tested also on uncorrupted images, which demonstrates its utility as a generic image reconstruction algorithm.

%%%%% Background %%%%%
\section{Background} \label{sec:background}
 %\todo[inline]{IO: Background Needs to be rewritten}

Deep learning has recently shown great promise in reconstruction  of highly undersampled MR acquisitions with convolutional neural networks (CNNs) \cite{Hyun2018,Qin2019,Schlemper2018}. For example, Schlemper et al. \cite{Schlemper2018} proposed to use a deep cascaded network to generate high quality images, and 
Hauptmann et al.~\cite{Hauptmann2018}  proposed to use a residual U-net  to reduce aliasing artefacts due to undersampling with the purpose of accelerating image acquisition.

For automatic correction of CMR, Lotjonen et al. \cite{Lotjonen2005} used reconstructed short-axis and long-axis slices to optimise the locations of the slices using mutual information as a similarity measure.
Estimating high quality images from corrupted (or under-sampled) k-space has been a well investigated subject in the literature \cite{Han2018}.  
The problem can be addressed either in the k-space domain or the image domain. 
One choice is to correct the k-space before applying the inverse Fourier transform (IFT) as proposed by Han et al.  \cite{Han2018}. 
A more common approach is to use the IFT on k-space and learn a mapping between the corrupted reconstructed images and good quality images. 
To this end, a variety of image denoising techniques can be utilized such as autoencoders \cite{Xie2012}, residual learning networks \cite{Zhang2017} or wide networks \cite{Liu2017}. Zhu et al. \cite{Zhu2018} proposed an end-to-end image reconstruction approach (Automap) for MR and evaluated it on undersampled k-space data.

\section{Methods}
\label{sec:method}

\begin{figure}[htb]
  \centering
  \centerline{\includegraphics[width=\linewidth]{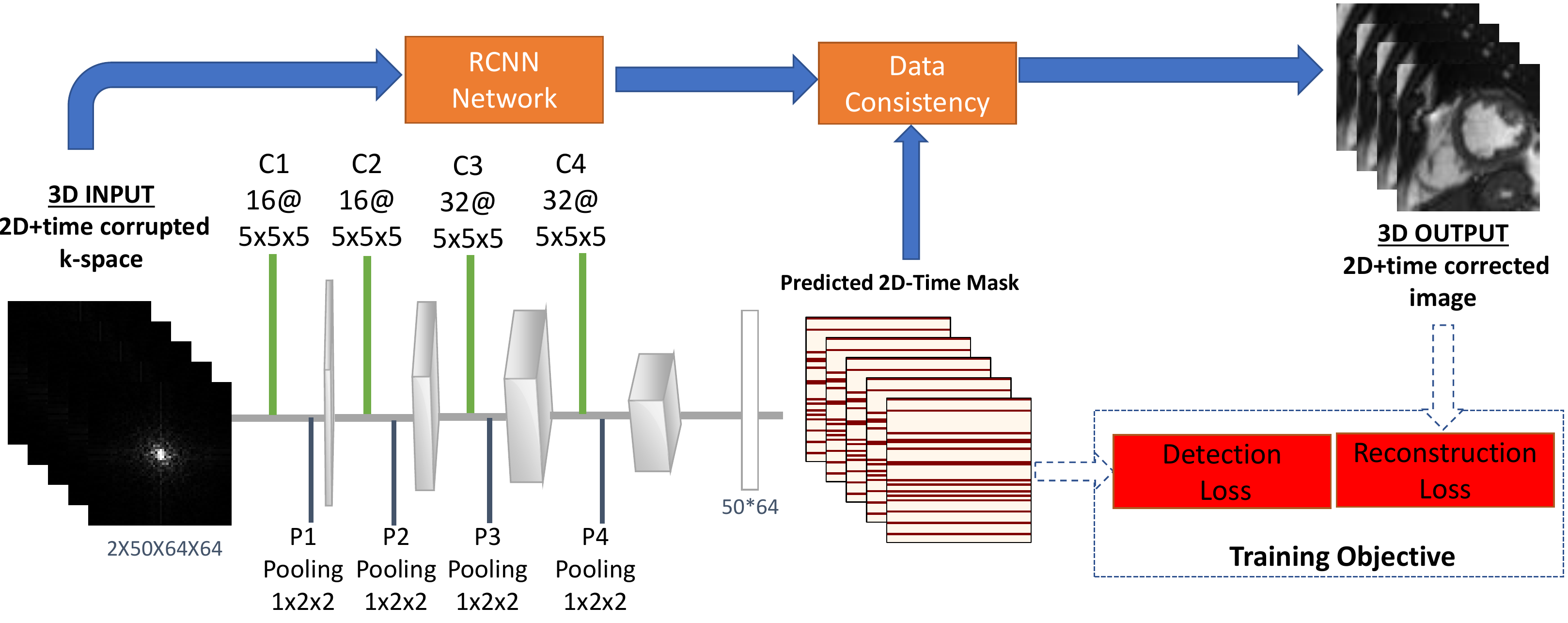}}
\caption{The CNN architecture for motion artefact correction. The proposed network architecture consists of two building blocks 1) A corrupted k-space line detection network to define the data-consistency term; 2) A recurrent neural network (RCNN) architecture to correct image artefacts.}
\label{fig:Model}
\end{figure}

Our network architecture consists of two sub-networks that are trained jointly as visualized in Fig. \ref{fig:Model}. The first network is an artefact detection network which is used to identify potentially corrupted k-space lines and hence define a data-consistency term. and the second network is a recurrent convolutional neural network (RCNN) used for reconstruction using this data-consistency term \cite{Qin2019}. Details of both networks are provided below.

\subsection{Network Architecture}

The proposed artefact detection CNN consists of eight layers 
The architecture of our network follows a similar architecture to \cite{Tran2017}, which was originally developed for video classification using a spatio-temporal 3D CNN. 
In our case we use the third dimension as the time component and use 2D+time mid-ventricular sequences as the input to the network. 
%The input CMR images are intensity-normalized and cropped to $64 \times 64$ pixels.
Each image sequence has 50 time frames. 
The network has 4 convolutional layers and 4 pooling layers, 1 fully-connected layer and a softmax loss layer to predict corrupted k-space lines.  
After each convolutional layer, a ReLU activation is used. 
We then apply pooling on each feature map to reduce the filter responses to a lower dimensionality. 
We apply dropout with a probability of 0.2 at all convolutional layers and after the first fully connected layer for regularization. 
All of these convolutional layers are applied with appropriate padding of 2 and stride of 1.

The reconstruction network features a RCNN architecture \cite{Qin2019}. This network reconstructs high quality cardiac MR images from highly undersampled k-space data by jointly exploiting the dependencies of the temporal sequences as well as the iterative nature of traditional optimisation algorithms. 
In addition, spatio-temporal dependencies are simultaneously learned by exploiting bi-directional recurrent hidden connections across time sequences. 
Any reconstruction network can replace this network in our architecture. We chose this particular network for its capability to incorporate information from different frames, which is instrumental in correcting  the artefacts that occur due to displacement of k-space lines in time.

%The networks consists of a bi-directional recurrent neural network for exploiting temporal information and convolutional recurrent layers to propagate information between iterations.\todo{BR: see two sentences before; two times the same message?} 

\subsection{Loss Function and Training}
\label{sect:aug}

Our training objective is a linearly weighted combination of the image reconstruction loss %(MSE loss)
and a cross-entropy loss for the detection of corrupted lines:
$$\mathcal{L}_{\text{total}}=   \lambda \mathcal{L}_{\text{detection}} +  (1-\lambda) \mathcal{L}_{\text{reconstruction}} $$
The reconstruction loss is computed using the mean square error, defined as: 
$$\mathcal{L}_{\text{reconstruction}}=  \dfrac{1}{N_{p}} \sum_{p=0}^{N_{p}}  (I_{x}(p)-I_{y}(p))^2   $$
where p denotes each pixel and $N_{p}$ denotes the total number of pixels in images $I_{x}$ and $I_{y}$. 
The detection loss is the cross entropy loss, defined as: 
$$\mathcal{L}_{\text{detection}}(pr,y)= \dfrac{1}{N_{l}} -(y \log(pr) + (1-y) \log(1-pr))$$
where $y$ is a binary indicator (0 or 1) indicating if a k-space line is corrupted or not and $pr$ is predicted probability of the line being uncorrupted.  $N_{p}$ denotes the total number of k-space lines in an image.

We used the Adam optimizer to minimise the binary cross entropy and mean square error loss function. 
$\lambda$, which defines the contribution of each loss was set to 0.3 using the validation set. 
The cross entropy term represents the dissimilarity of the predicted output distribution to the true distribution of labels after a softmax layer.  
The detection and reconstruction networks were pre-trained for 50 epochs separately to enable faster convergence. 
End-to-end training ended when the network did not significantly improve its performance on the validation set for a predefined number of epochs (100). 
An improvement was considered sufficient if the relative increase in performance was at least 0.5\%. 

During training, a batch-size of 50 2D+time sequences was used.  
The momentum of the optimizer was set to 0.90 and the learning rate was %0.0001. JC
$10^{-4}$.
The parameters of the convolutional and fully-connected layers were initialised randomly from a zero-mean Gaussian distribution. 
In each trial, training was continued until the network converged. 
Convergence was defined as a state in which no substantial progress was observed in the training loss. We used Pytorch for implementation of the network and training took around 3 hours on a NVIDIA Quadro P6000 GPU. 
After training, deployment of the network to correct a single image sequence took less than 1s.

%\vspace{-0.2cm}
\section{Experimental Results}
\label{sec:results}
%\vspace{-0.2cm}

We evaluated our algorithm on a subset of the UK Biobank dataset containing 300 datasets each consisting of 50
2D+time good quality cine CMR acquisitions at a mid-ventricular short axis slice. From each subject, the 50 temporal frames were used to generate synthetic motion artefacts. We used 200 datasets for training, 50 for validation and 50 for testing.
The total of 300 subjects
were chosen to be free of other types of image quality issues, such as missing axial slices, and were visually verified by an expert cardiologist for sufficient image quality. The details of the acquisition protocol of the UK Biobank dataset can be found in \cite{Petersen2016}. 

%Data prepration
\textbf{Data preprocessing:} Given a 2D+time cine CMR sequence of images we first normalise the pixel values between 0 and 1. Since the image dimensions vary from subject to subject, instead of image reshaping we use a motion information based ROI extraction to $64\times 64$ pixels 
%Avoiding image resampling is of particular importance for image quality assessment, because resampling can influence the image quality significantly due to interpolation.  The ROI is determined using an unsupervised technique based on motion as proposed in
\cite{Korshunova2016}. Briefly, the ROI is determined by performing a Fourier analysis of the sequences of images, followed by a circular Hough transform to highlight the center of periodically moving structures. 

%.rueckert: Is that true for UKBB. IO : yes it is true
\textbf{K-space corruption for synthetic data:} We generated k-space corrupted data in order to simulate motion artefacts. We followed a Cartesian sampling strategy for k-space corruption to generate synthetic but realistic motion artefacts \cite{Oksuz2018},~\cite{Oksuz2019}. 
We retrospectively transformed each 2D short axis sequence to the Fourier domain and changed a number (0,2,4,8,16) of Cartesian sampling k-space lines to the corresponding lines from other cardiac phases to mimic motion artefacts. These lines were selected randomly in order to mimic the randomness of real motion artefacts.
The  newly introduced k-space lines were also selected randomly from all other frames in the image sequence. 
In this way CMR images with artefacts were generated from the original `good quality' images in the training set. This is a realistic approach as the motion artefacts that occur from mis-triggering  arise from similar displacement (in time) of an arbitrary set of k-space lines.

%Methods of comparison
\textbf{Methods of comparison:} We compared our algorithm to a variety of classes of artefact correction strategy
%to produce robust artefact correction. 
%The methods of comparison cover all possible combinations of motion artefact correction in the image domain
as outlined in Section \ref{sec:background}. 
For image-to-image to artefact removal (i.e. post-reconstruction), we used a deep network based on residual learning (DNCNN) as well as a wide network with larger receptive fields and more channels in each layer as proposed in \cite{Liu2017} (WIN5). 
We also compared our method to a reconstruction algorithm that uses an end-to-end correction methodology \cite{Oksuz2018a} (Automap-GAN) based on Automap \cite{Zhu2018}. Additionally, we compared our approach to its variants: 
1) training detection and reconstruction networks separately (Proposed-separate); and
2) considering the corruption mask as a pre-determined mask to illustrate the top performance achievable in this setup  (Proposed-known mask).

% Table
\begin{table} [tb]
\centering
\caption{Mean image quality results of image quality correction for motion artefacts for corrupted and uncorrupted inputs. Uncorrupted results use the correct k-space as input and indicates the potential of our method to be used as a global image reconstruction framework.}
\begin{tabular}{lcccccc}
\hline
& \multicolumn{3}{c}{Corrupted} & \multicolumn{3}{c}{Uncorrupted}  \\
\cline{2-4}
\cline{5-7}
Methods                                 & PSNR & RMSE  &SSIM   & PSNR & RMSE  &SSIM \\
\hline 

Baseline                                & 26.3  & 0.068 & 0.821 & -     & -     & - \\
DNCNN  \cite{Zhang2017}                 & 30.8  & 0.049 & 0.845 & 36.7  & 0.005 & 0.905\\
Win5    \cite{Liu2017}                 & 32.2  & 0.041 & 0.853 & 37.2  & 0.004 & 0.913 \\
Automap-GAN   \cite{Oksuz2018a}         & 34.8  & 0.028 & 0.878 & 38.7  & 0.003 & 0.927 \\
\hline
Proposed-separate                       & 34.7  & 0.026 & 0.879 & 39.3  & 0.003 & 0.947 \\
\textbf{Proposed-end to end}    & \textbf{37.1} & \textbf{0.023}& \textbf{0.890} & \textbf{40.8}  & \textbf{0.002} & \textbf{0.972}  \\
Proposed-known Mask                     & 38.9  & 0.019 & 0.901  &  - & -& - \\
\hline
\end{tabular}
\label{table:1}
\end{table}

\textbf{Quantitative results:} Table \ref{table:1} shows the image quality metrics for the corrected images produced by each image artefact correction algorithm for corrupted and original images. 
For these experiments the ground truth was the uncorrupted original 2D+time image sequence and we use peak signal to noise ratio (PSNR), root mean square error (RMSE) and structural similarity index metric (SSIM) for evaluation. 
The proposed end-to-end k-space detection and correction algorithm outperforms the other methods. As can be appreciated, the joint end-to-end network performs better compared to separate training of both architectures. Compared to the image-to-image denoising techniques (Win5, DNCNN), k-space based correction (Automap-GAN) results in improved reconstructions of the images. We have also shown results on using original images as input to illustrate the capability of our method as a general image reconstruction algorithm. Our method outperforms the other state of the art techniques and does not diminish the image quality of the original k-space thanks to the detection network. Baseline and proposed-known mask methods provide perfect image quality in case of uncorrupted images and therefore omitted.

% Figure
 \begin{figure}[htb]
\centering
  \centerline{\includegraphics[width=\linewidth]{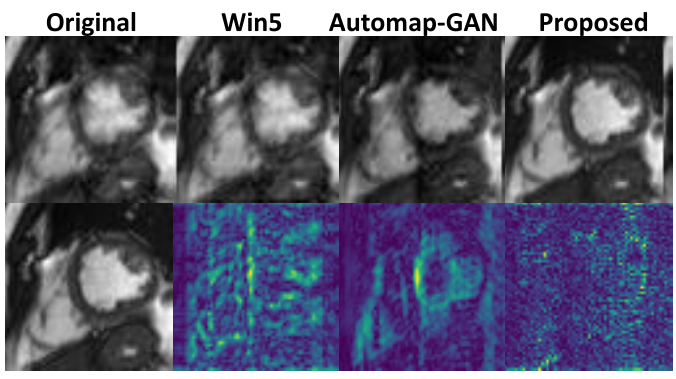}}
\caption{The results produced by Win5 (second column), Automap-GAN (third column)  and proposed  algorithm (final column). In the first column, the top row shows the corrupted image and the bottom row shows the corresponding uncorrupted image. It is evident from the difference images in the bottom row that image quality is recovered at the septum using proposed method.}
\label{fig:Result}
\end{figure}

\textbf{Qualitative results:} In Figure \ref{fig:Result} we illustrate the performance of our technique on artefact correction in comparison to the top state-of-the-art techniques \cite{Oksuz2018a,Liu2017}. 
The difference image shows improved image quality with the proposed technique, especially in the left ventricular and right ventricular regions and with regard to the sharpness of the myocardial boundaries. These results demonstrate that the network reduced the impact of k-space corruption on reconstruction quality, as the (beating) ventricles and their myocardial borders are mostly affected by such corruption.

%\todo[color=pink!40]{EP: Maybe add some arrows to show differ-ences between Automap-GAN and your results, for example it seems that the myocardium is better delinated with your method, specially on the septum}
%\todo[color=pink!40]{EP: It could be interesting to show an example of a corrupted image form the database and show how your method improves it and now can be used to compute volumes or sth like that'}

\section{Discussion and Conclusion}
\label{sec:discussion}

In this paper, we have proposed a CNN-based technique for correcting motion-related artefacts in a large 2D+time CINE CMR dataset.
%with high PSNR and SSIM. 
We have addressed the issue of incorrect k-space lines using a combined architecture to detect, correct and reconstruct images. The proposed network clearly outperforms competing algorithms. Moreover, the current architecture can also be used as a global image reconstructor, as we have shown that it does not diminish the quality of uncorrupted images, compared to the original reconstruction performed on the scanner.
%\todo{AK: Did I miss this? Where have you shown that it doesn't diminish the quality of uncorrupted images?}

We have shown for the first time that a 3D CNN based neural network architecture is capable of classifying k-space lines that cause motion artefacts.  
%To the best knowledge of the authors, this is the first paper that has proposed an image reconstruction network for under-sampled k-space to address image artefacts. 
Our work brings fully automated assessment of ventricular function from CMR imaging a step closer to clinically acceptable standards, enabling reconstruction of high quality images from data containing artefacts in order to enable their analysis in large imaging datasets such as the UK Biobank. 
In future work, we plan to validate our method on the entire UK Biobank cohort, which is eventually expected to be 100,000 CMR images.
%\todo{BR: if space maybe add a line about atacking other artefacts, such as those originating from metal implants?} 

\textbf{Acknowledgments}\\
This work was supported by an EPSRC programme Grant (EP/P001009/1) and the Wellcome EPSRC Centre for Medical Engineering at the School of Biomedical Engineering and Imaging Sciences, King’s College London (WT 203148/Z/16/Z). This research has been conducted using the UK Biobank Resource under Application Number 17806. The GPU used in this research was generously donated by the NVIDIA Corporation.

\end{document}